\documentclass[
aps,%
12pt,%
final,%
notitlepage,%
oneside,%
onecolumn,%
nobibnotes,%
nofootinbib,%
superscriptaddress,%
noshowpacs,%
centertags]%
{revtex4}

\usepackage{epsf}

\usepackage{amsmath}
\usepackage{graphicx}
\usepackage[cp866]{inputenc}
\usepackage[english,russian]{babel}

\newcommand{\be}{\begin{equation}}
\newcommand{\ee}{\end{equation}}
\newcommand{\bea}{\begin{equation}\begin{aligned}}
\newcommand{\eea}{\end{aligned}\end{equation}}

\newcommand{\beq}{\begin{eqnarray}}
 \newcommand{\eeq}{\end{eqnarray}}

\def\fun#1#2{\lower3.6pt\vbox{\baselineskip0pt\lineskip.9pt
\ialign{$\mathsurround=0pt#1\hfil ##\hfil$\crcr#2\crcr\sim\crcr}}}

\newcommand{{\SD}}{\rm SD}

\newcommand{\lan}{\langle}
\newcommand{\ran}{\rangle}
\begin{document}

%\selectlanguage{russian}

\title{Low-energy relation for the trace of the energy-momentum tensor in QCD and the gluon condensate in a magnetic field}
\author{N.O. Agasian}
\affiliation{
Institute of Theoretical and Experimental Physics,\\
117218 Moscow, Russia}
\affiliation{
National Research Nuclear University MIFI, \\
115409 Moscow, Russia}

\begin{abstract}
A treatment is given of the nonperturbative QCD vacuum in a magnetic field.
The low-energy equation for the trace of the energy-momentum tensor in a magnetic field is derived.
It is shown that the derivatives with respect to a magnetic
field of the quark and gluon contributions to the trace of the energy-momentum tensor are equal.
The dependence of the gluon condensate on the magnetic field strength is derived both for strong and weak fields.
\end{abstract}

\maketitle

1. The phase structure of vacuum in an external
magnetic field $B$ became an important object of investigations.
It has been recently shown that strong magnetic
fields with strengths $eB\sim 10^2 \div 10^4$ МэВ$^2$ can be
generated in heavy-ion collisions.
These magnetic fields can initiate the observed phenomena (so-called
chiral magnetic effect) \cite{Kharzeev:1998kz, Kharzeev:2004ey, Kharzeev:2007jp, Fukushima:2008xe, Fukushima:2010fe}
in experiments at the RHIC and LHC.
Strong magnetic fields $eB\sim \Lambda_{QCD}^2$ could have existed in the early Universe at the energy
scale of strong interactions. These magnetic fields can lead to new interesting phenomena at the stage of the
quark-hadron phase transition \cite{Shushpanov:1997sf, Agasian:1999sx, Agasian:2001hv, Agasian:2001ym,
Agasian:2008tb, Fraga:2008qn, Mizher:2010zb, Fraga:2012rr, Agasian:2008zz, Gatto:2010pt, Gatto:2012sp, Ayala:2015lta,
Ayala:2015qwa, Andersen:2012bq, Andersen:2014xxa, Menezes:2008qt, Menezes:2009uc, Bali:2012zg,
Orlovsky:2013aya, Orlovsky:2013wjd, Andreichikov:2013zba, Miransky:2015ava, Ilgenfritz:2012fw, Ilgenfritz:2013ara,
Buividovich:2009wi, D'Elia:2010nq}.

Quantum-field theories involve important relations
following from the symmetry properties of a theory.
Search for symmetries and constraints imposed by
them on the physical characteristics of the system are
of particular importance in QCD, which is a theory
with confinement, where ``observables'' are composite
states-hadrons.
Low-energy theorems or Ward identities
(scale and chiral) are of fundamental importance
for a deeper insight into the nonperturbative properties
of the QCD vacuum.
Low-energy theorems of QCD
were obtained in the early 1980s \cite{Novikov}.  Low-energy theorems
of QCD, which follow from the general symmetry
properties and are independent of the details of the
confinement mechanism, make it possible to obtain
information sometimes otherwise inaccessible.
In addition, they can be used as ``physically reasonable''  constraints when developing effective theories and
various models of the QCD vacuum.
The low-energy theorems of QCD at $T\neq0$, $\mu_q \neq 0$ were obtained in \cite{Ellis:1998kj, Shushpanov:1998ce}.
Low-energy theorems in a magnetic field and
its applications to various physical processes were
studied in \cite{Agasian:2011st, Agasian:2011sm, Agasian:2003yw, Agasian:2000hw}.
Low-energy relations for the
energy-momentum tensor at finite temperature were
analyzed in
\cite{Agasian:2001sv, Agasian:2001bj}.

In this work, a low-energy relation for the trace of the energy-momentum tensor in QCD in a magnetic field is
obtained. This relation is used to find the magnetic field
dependence of the gluon condensate in the limiting
case of strong and weak fields.

2. In the Euclidean formulation, the partition
function of QCD in the presence of an external Abelian
field $A_\mu$ can be represented in the form
\be
 Z=\exp \left \{ -\frac{1}{4e^2}
\int d^4x F^2_{\mu\nu} \right \}
\int[DC][D\bar q][Dq]
\exp \left \{ -\int d^4x {\cal L} \right \},
 \label{eq_1}
  \ee
where the QCD Lagrangian in a background field has
the form
  \be
 {\cal L}=\frac{1}{4g^2_0}
 (G^a_{\mu\nu})^2
 + \sum_{q=u,d} \bar q[\gamma_\mu
 (\partial_\mu-iQ_q A_\mu-i\frac{\lambda^a}{2} C^a_\mu)+m_{0q}]q.
 \label{eq_2}
  \ee
 Here, $Q_q$ -- is the charge matrix for quarks with flavor
 $q=(u,d)$ and the bare mass $m_{0q}$ and the ghost and
gauge-fixing terms are not written explicitly for simplicity.
The energy density is given by the expression $V_4\varepsilon$
$(B,m_{0u},m_{0d})=-\ln Z$.
The relation for the gluon
condensate ($\langle G^2\rangle\equiv \langle (G^a_{\mu\nu})^2\rangle$) follows from Eq.(\ref{eq_1}):
  \be
    \langle G^2\rangle (B,m_{0u},m_{0d})=4\frac{\partial \varepsilon}{\partial(1/g^2_0)}~.
 \label{4}
 \ee
The system described by the partition function given
by Eq. (\ref{eq_1}) is characterized by the set of dimensional
parameters $M, B, m_{0q}(M)$ and the dimensionless
charge $g^2_0(M)$, where $M$ is the ultraviolet-cutoff mass.
 At the same time, it is possible to consider the renormalized
energy density
 $\varepsilon_R$ and, with the use of dimensional
and renormalization group properties
$\varepsilon_R$, to
represent Eq. (\ref{4}) in the form including the derivatives
with respect to both the physical parameter $B$ and
renormalized masses $m_q$.

The phenomenon of dimensional transmutation
leads to the appearance of the nonperturbative dimensional
parameter
 \be
  \Lambda= M \exp
 \left \{ \int^\infty_{\alpha_s(M)}\frac{d\alpha_s}{\beta(\alpha_s)}
 \right \}~,
  \label{5}
  \ee
where
  $\alpha_s=g^2_0/4\pi$ and $\beta(\alpha_s)=d\alpha_s(M)/d
  \ln M$ is the Gell-Mann--Low function.
It is well known that the quark mass has an anomalous dimension and depends on the scale $M$.
The renormalization group
equation for the running mass $m_0(M)$ has the form $d\ln m_0/d\ln
M=-\gamma_m$ and the $\overline{MS} $ scheme is used,
where
$\beta$ and $\gamma_m$ are independent of the quark mass \cite{Shushpanov:1998ce,muta}.
The renormalization group invariant mass has the form
\be
m_q=m_{oq}(M)\exp\{\int^{\alpha_s(M)}\frac{\gamma_{m_q}(\alpha_s)}{\beta(\alpha_s)}
d\alpha_s\}~,
 \label{6}
 \ee

 Since the energy density is a renormalization group
invariant quantity, its anomalous dimension is zero. Therefore, $\varepsilon_R$
has only a normal (canonical) dimension equal to 4.
In view of the renormalization group invariance of the quantity
of $\Lambda$,  $\varepsilon_R$ can be written in the most general form
\be
\varepsilon_R=\Lambda^4 f(\frac{B}{\Lambda^2}, \frac{m_u}{\Lambda},
\frac{m_d}{\Lambda})~,
 \label{7}
 \ee
where $f$ is a certain function.
It follows from Eqs.(\ref{5}),(\ref{6}) and (\ref{7}) that
\be
\frac{\partial \varepsilon_R}{\partial(1/g^2_0)}=
 \frac{\partial
\varepsilon_R}{\partial\Lambda} \frac{\partial\Lambda}{\partial(1/g^2_0)} +
\sum_q \frac{\partial \varepsilon_R}{\partial m_q} \frac{\partial
m_q}{\partial(1/g^2_0)}~,
 \label{8}
 \ee

\be \frac{\partial m_q}{\partial(1/g^2_0)}=-4\pi\alpha^2_s
m_q\frac{\gamma_{m_q}(\alpha_s)}{\beta(\alpha_s)}~.
 \label{9}
 \ee
In view of (\ref{4}) the gluon condensate is given by the expression
\be
\lan G^2\ran (B, m_u, m_d)=
\frac{16\pi\alpha_s^2}{\beta(\alpha_s)}(4-2B\frac{\partial}{\partial
B}-\sum_q(1+\gamma_{m_q})m_q\frac{\partial}{\partial {m_q}}) \varepsilon_R.
\label{10}
 \ee
It is convenient to choose the scale large enough to
take the lowest order in the expansion of the Gell-
Mann-Low function,
   $\beta(\alpha_s)\to -
b\alpha^2_s/2\pi$, where $b=(11 N_c-2N_f)/3$ and $1+\gamma_m\to
1$.
Thus, the equations for condensates have the form \cite{Agasian:2001hv}
\be
\lan G^2\ran (B)
=-\frac{32\pi^2}{b} (4-2B\frac{\partial}{\partial
B}-\sum_q m_q\frac{\partial}{\partial m_q}) \varepsilon_R\equiv
 -\hat D\varepsilon_R~,
 \label{11}
 \ee
 \be
 \lan\bar q q\ran (B)=\frac{\partial \varepsilon_R}{\partial {m_q}}~.
 \label{12}
 \ee

 3. In QCD, the effective energy density from which
condensates
$\lan G^2\ran(B)$   and  $\lan \bar q q\ran(B)$ can be obtained by
Eqs. (\ref{11}) and (\ref{12}), respectively, has the form
\be
\varepsilon_{eff}(B)=\varepsilon_{vac}+\varepsilon_h(B),
\label{14}
\ee
where
$\varepsilon_{vac}=\frac14\lan\theta_{\mu\mu}\ran$ is the nonperturbative vacuum
energy density at $B=0$ and
\be
\lan
\theta_{\mu\mu}\ran=-\frac{b}{32\pi^2} \lan G^2\ran+\sum_{q=u,d}
m_q\lan\bar qq \ran
\label{15}
\ee
is the trace of the energy-momentum tensor.
In Eq. (\ref{14}), $\varepsilon_h(B)$ is the energy density created by hadrons
in the magnetic field.
The quark and gluon condensates are given by the formulas
\be
\lan \bar qq\ran
(B)=\frac{\partial \varepsilon_{eff}}{\partial m_q},
\label{17}
\ee
\be
\lan G^2\ran (B)= -\hat D\varepsilon_{eff},
\label{18}
\ee
where the operator $\hat D$, according to Eq. (\ref{11}), has the form
\be \hat
D=\frac{32\pi^2}{b} (4-2B\frac{\partial}{\partial B}-\sum_q
m_q\frac{\partial}{\partial m_q})~.
\label{19}
\ee

Below, the case $B=0$
is considered with the use of the
low-energy theorem for the derivative of the gluon
condensate with respect to the quark mass \cite{Novikov}
\be
\frac{\partial}{\partial m_q}\lan G^2\ran= \int d^4 x\lan G^2(0)
\bar q q(x)\ran =-\frac{96\pi^2}{b}\lan \bar q q\ran+O(m_q),
\label{20}
\ee
where
$O(m_q)$ stands for terms linear in the masses of
light quarks. The resulting relation has the form
\be
\frac{\partial\varepsilon_{vac}}{\partial m_q}=-
\frac{b}{128\pi^2}\frac{\partial}{\partial m_q} \lan
G^2\ran+\frac{1}{4}\lan \bar q q\ran =\frac34 \lan \bar q
q\ran+\frac14\lan \bar q q\ran=\lan \bar q q\ran.
\label{21}
\ee
Three-fourths of the quark condensate originates from
the gluon part of the nonperturbative energy density of
vacuum. The following expression for the gluon condensate is obtained similarly
\be
-\hat D\varepsilon_{vac}=\lan G^2\ran.
\label{22}
\ee

In order to obtain the dependence of the quark and
gluon condensates on the magnetic field
$B$ , it is convenient
to use the Gell-Mann--Oakes--Renner relation ($\Sigma=|\lan\bar u u\ran|=|\lan\bar dd\ran|$)
\be F^2_\pi M^2_\pi=-\frac12(m_u+m_d)\lan \bar
uu+\bar dd\ran=(m_u+m_d)\Sigma~.
\label{23}
\ee
Then, the following relations are derived:
\be
\frac{\partial}{\partial
m_q}=\frac{\Sigma}{F^2_\pi} \frac{\partial}{\partial M^2_\pi}~,
\label{24}
\ee
\be \sum_qm_q\frac{\partial}{\partial
m_q}=(m_u+m_d)\frac{\Sigma}{F^2_\pi}\frac{\partial}{\partial
M^2_\pi}=M^2_\pi\frac{\partial}{\partial M^2_\pi}~,
\label{25}
\ee
\be \hat D=\frac{32\pi^2}{b}(4-2B\frac{\partial}{\partial
B}-M^2_\pi\frac{\partial}{\partial M^2_\pi})~.
\label{26}
\ee

Within the approach described above, it is possible
to obtain a low-energy relation for the quantum anomaly
in the trace of the energy-momentum tensor in the
magnetic field.
The leading contribution to the energy
density in the magnetic field comes from the lightest
hadrons, namely, $\pi$-mesons.
The general expression
for the energy density in the magnetic field has the form
 \be
 \varepsilon_\pi=B^2\varphi(M_\pi^2/B)~,
 \label{36}
 \ee
where $\varphi$ is a function of the ratio $M_\pi^2/B$.
Then, we obtain the relation
 \be
 (4-2B\frac{\partial}{\partial B}-M^2_\pi\frac{\partial}{\partial
 M^2_\pi}) \varepsilon_\pi=M^2_\pi\frac{\partial \varepsilon_\pi}{\partial M^2_\pi}~.
 \label{37}
 \ee
Taking into account Eqs. (\ref{17},\ref{18}), (\ref{21},{22}) and (\ref{37}),
we arrive at the relations
\be
\Delta \lan \bar qq\ran =\frac{\partial \varepsilon_\pi}{\partial m_q},~~
\Delta \lan G^2\ran=-\frac{32\pi^2}{b} M^2_\pi\frac{\partial
\varepsilon_\pi}{\partial M^2_\pi}~,
\label{38}
\ee
where $ \Delta \lan \bar
qq\ran= \lan \bar qq\ran_B- \lan \bar qq\ran $ and $\Delta \lan
G^2\ran=  \lan G^2\ran_B- \lan G^2\ran.$
In view of Eq. (\ref{25}), Eq.(\ref{38}) can be written in the form
\be
\Delta \lan G^2\ran=-\frac{32\pi^2}{b} \sum_q m_q\Delta  \lan \bar qq\ran~.
\label{39}
\ee
The division of both sides of Eq. (\ref{39}) by
$\Delta B$ and the
passage to the limit $\Delta B\to 0$ give
\be
\frac{\partial \lan G^2\ran}{\partial B}=-\frac{32\pi^2}{b} \sum_q
m_q\frac{\partial\lan \bar qq\ran}{\partial B}~.
\label{40}
\ee
This relation can be rewritten in the form
\be
\frac{\partial \lan \theta^g_{\mu\mu}\ran}{\partial
B}=\frac{\partial \lan \theta^q_{\mu\mu}\ran}{\partial B}~,
\label{41}
\ee
where $\lan \theta^q_{\mu\mu}\ran=\sum m_q \lan
\bar qq\ran$ and $\lan
\theta^g_{\mu\mu}\ran=(\beta(\alpha_s)/16\pi\alpha^2_s) \lan
G^2\ran$ are the quark and gluon contributions, respectively,
to the trace of the energy-momentum tensor.
This result was obtained with the use of the Gell-Mann--Oakes--Renner relation; consequently,
(\ref{40}) and (\ref{41}) are valid in the theory with light quarks.
Therefore, in the region where excitations of
massive hadrons, as well as the interaction of pions,
can be neglected, Eq.
(\ref{41}) becomes an exact theorem of QCD.

In the chiral limit ($m_q=0$), it follows from Eq. (\ref{39}) that the gluon condensate (in contrast to the quark
condensate) is independent of the magnetic field ($\Delta \lan G^2\ran=0$). This is because massless noninteracting
$\pi$-mesons constitute a scale-invariant system and,
hence, do not contribute to the trace of the energy-momentum tensor.

4. The above relations in combination with the
magnetic-field dependence of the quark condensate
allow the determination of the magnetic-field dependence
of the gluon condensate.
In chiral perturbation
theory, the power-series expansion of the quark condensate
in the magnetic field \cite{Agasian:1999sx} is performed in the
parameter $\xi=eB/(4\pi F_\pi)^2$. Then, within chiral perturbation
theory, the field at $\xi<1$ can be treated as weak.
However, from the physical point of view, it is
necessary to consider the radius of curvature of the
trajectory of a particle in the field (Larmor radius)
$1/\sqrt{eB}$. If the Larmor radius is smaller than the charge
radius of the $\pi$-meson, $1/\sqrt{eB}<r_\pi$,
the particle cannot
be considered as a point particle and the quark structure
of the $\pi$-meson should be taken into account. For
weak magnetic fields, the shift of the quark condensate
in the magnetic field was found in \cite{Agasian:2001hv} ($\Delta \Sigma=-\Delta  \lan \bar qq\ran$)
 \be
   \frac{\Delta \Sigma}{\Sigma}=
   \frac{(eB)^2}{96\pi^2F^2_\pi M^2_\pi}~.
   \label{42}
   \ee
According to Eq. (\ref{39}) and the Gell-Mann--Oakes--Renner relation given by Eq. (\ref{23}), the behavior of the
gluon condensate in a weak magnetic field is described by the expression
\be
\lan G^2\ran (B)=\lan G^2\ran+ \frac{(eB)^2}{3b}~.
\label{43}
\ee
In the case of a strong magnetic field, the quark condensate
was calculated in \cite{Shushpanov:1997sf} in the form
\be
 \frac{\Delta \Sigma}{\Sigma}=
\frac{eB}{(4\pi F_\pi)^2} \ln 2~.
\label{44}
\ee
Then, the gluon condensate in strong fields is given by
the expression
\be
\lan G^2\ran (B)=\lan G^2\ran+ \frac{2\ln 2}{b}M^2_\pi eB~.
\label{43}
\ee

A quadratic increase in the gluon condensate in
weak magnetic fields is in agreement with lattice calculations \cite{D'Elia:2015dxa},
where this phenomenon was called gluon catalysis.
%Также квадратичный рост глюонного конденсата в слабых полях был получен в рамках правил сумм КХД в работе
%\cite{Ayala:2015qwa}.

5. The nonperturbative QCD vacuum in a magnetic
field has been studied. A low-energy relation for
the trace of the energy-momentum tensor in the magnetic
field has been obtained ab initio. Analytical
expressions for the gluon condensate in the magnetic
field have been derived in the limits of strong and weak fields.

%-----------------------------------------------------------------------------

\end{document}